# Polyply: a python suite for facilitating simulations of (bio-) macromolecules and nanomaterials

Fabian Grünewald[1], Riccardo Alessandri[1,2], Peter C. Kroon[1], Luca Monticelli[3], Paulo C.T. Souza[3], Siewert J. Marrink[1]

[1] Groningen Biomolecular Sciences and Biotechnology Institute and Zernike Institute for Advanced Materials, University of Groningen, Groningen, The Netherlands

[2] Pritzker School of Molecular Engineering, University of Chicago, Chicago, IL 60637, USA

[3] Molecular Microbiology and Structural Biochemistry, UMR 5086 CNRS and University of Lyon, Lyon, France

**ABSTRACT:** Molecular dynamics simulations play an increasingly important role in the rational design of (nano)-materials and in the study of biomacromolecules. However, generating input files and realistic starting coordinates for these simulations is a major bottleneck, especially for high throughput protocols and for complex multi-component systems. To eliminate this bottleneck, we present the polyply software suite that leverages 1) a multi-scale graph matching algorithm designed to generate parameters quickly and for arbitrarily complex polymeric topologies, and 2) a generic multi-scale random walk protocol capable of setting up complex systems efficiently and independent of the target force-field or model resolution. We benchmark quality and performance of the approach by creating melt simulations of six different polymers using two force-fields with different resolution. We further demonstrate the power of our approach by setting up a multi lamellar microphase-separated block copolymer system for next generation batteries, and by generating a liquid-liquid phase separated polyethylene oxide-dextran system inside a lipid vesicle, featuring both branching and molecular weight distribution of the dextran component.

## 1 – Introduction

Molecular dynamics (MD) simulations of (bio-)macromolecules have become a powerful tool for researchers to complement experimental assays. Whereas simulations of single polymers melts or mixtures have been used since the advent of modern MD[1,2], recently the trend goes towards studying more complex multicomponent systems either of purely synthetic materials or biologically synthetic hybrid macromolecules.[3–9] Examples of such systems range from polyelectrolyte complex coacervates[10] to next generation polymer batteries.[11,12] Whereas simulations of these complex systems are typically focused on studying the self-assembly or understanding structure-function relationships, much effort is now being directed towards developing MD based protocols for virtual high throughput (HT) screening of polymers as exemplified by the material genome initiative.[13–16] HT screening of polymers by MD is expected to complement experimental HT approaches, because it is typically less costly than synthetic exploration and gives access to properties not easily accessible by purely experimental HT approaches. Such combined approaches will enable researchers to survey a larger combinatorial space and filter possible candidates more efficiently.[14,16] Applications of such procedures range from design of novel antimicrobial polymers to biodegradable polymers.[14]

Though the avenue of HT simulations is promising, it requires programs to build topologies and simulation boxes in a quick, reliable and consistent manner. Moreover, given the hierarchy of spatiotemporal scales underlying the behavior of polymer-based systems, models with both all-atom and coarse-grained (CG) resolution are required. While a wide range of programs[17–26] are available for MD simulations of biologically relevant systems such as proteins, lipid membranes and DNA, the support for simulation of synthetic and biosynthetic hybrid macromolecules is largely lacking. To our knowledge there are no programs which can generate both input parameters and coordinates for arbitrarily complex polymeric systems independent of the force-field and compatible with HT approaches. Depending on the molecule or system there are a number of specific solutions.[20,27–30] Some of those are in principle capable of generating parameters and coordinates.[29,30] However, available programs typically support only one force-field and are limited to specific (mostly linear) polymers implemented by the developers. Website implementations[28,30] have the added problem that performance relies on server-load and it involves human-time having to interface with the website. In addition, coordinates for more complex systems such as micro phase separated polymers and hybrid nanoparticle blends are frequently generated by (multi-scale) self-assembly[31–33] or custom in-house building scripts[34,35].

The general lack of programs supporting all-atom and CG polymer simulations limits the use of MD simulations for both large versatile systems and high throughput research of (bio-) macromolecular systems. This is especially true for non-experts in the field. To this end we identify five major challenges that need to be overcome. 1) The program needs to be able to generate both coordinates and parameters, resolution and force-field independent. Accurate CG models are often based on atomistic polymers, thus modeling both of those is an integral part of HT model development. In addition, the program needs to be force-field agnostic, as some force-fields work better for specific polymers than others. 2) There needs to be an easy-to-use pipeline for generating input files and coordinates based on the system composition. Input files should be generated from sequences of arbitrarily complex polymers, including different degrees of branching and statistical distribution of residues along the chain. 3) The program needs to be able to combine input parameters and coordinates of polymeric systems with a variety of biomolecular structures, like proteins, lipid bilayers and nucleotides. For example, manipulation of proteins and other biomolecules by polymer grafting is an integral part of enhancement strategies.[36] 4) It needs to be capable of setting up complex systems without the need of extensive relaxation. Polymer melts, blends with nano-particles and phase separated systems are highly important in material science and when studying bio-synthetic hybrid molecules, and closely capturing their heterogeneity in the starting structures saves valuable computer resources. 5) Generation of both the coordinate and parameter files needs to be fast enough to enable HT research.

Here we present the open-source polyply software suite which addresses the five major challenges presented above. It facilitates the generation of input parameters and coordinates for MD simulations of (bio-)macromolecules and nanomaterials. Using a graph-based algorithm, polyply allows users to generate parameter files of arbitrarily composed and branched polymers for any force-field from simple library files and the residue graph. A residue graph contains the sequence of residues of the polymer, but in addition it also records which residues are connected. Using a multiscale random walk, polyply can also be utilized to generate starting coordinates for any force-field and at any target resolution. This includes complex arrangements like micro phase separated polymeric systems or multicomponent polymer solutions enclosed in lipid vesicles. To maximize accessibility of the models and code, polyply is distributed as via the python package index. Furthermore, polyply is developed using modern software development practices (such as code review, continuous integration testing, and semantic versioning) as outlined in the recent whitepaper by the BioExcel consortium.[37] These practices ensure both the integrity of the code and the library data files.

The remainder of this paper is organized as follows. First, we present the algorithms behind he input parameter generation and the coordinate generation. Subsequently we show the capabilities of our program based on three examples from pure materials science to bio-molecular science, some at the atomistic level based on the GROMOS[38] force-field, others at the CG level using the Martini[39,40] force-field. They exemplify the capabilities of polyply to be compatible with HT approaches, generate parameters force-field independently, and set up complex systems. Finally we discuss the limitations of our approach and sketch possible future directions for its further development.

## 2 – Algorithms and code

The general code design accounts for the fact that generation of input parameter files and coordinate generation are in principle separate problems. However, both problems can make use of the same infrastructure, which is centered around exploiting graph representations of molecules. Thus, the polyply software suite consist of separate lone standing programs, which utilize the same libraries. At the moment two programs, 'gen_params' and 'gen_coords', are available for input parameter generation and coordinate generation, respectively. In addition, an auxiliary program for sequence generation is provided, which is further discussed in the supporting information. In the following two sections we explain the algorithm and ideas behind parameter file generation and system coordinate generation.

## 2.1 – Parameter file generation

The problem of generating parameter files is treated as a graph transformation within polyply. Graph transformations are commonly used in other tools for generation input parameters as well.[19,29] The graph transformation in polyply takes a residue graph and maps it into a higher resolution graph, which is agnostic to the target resolution. A graph consists of nodes and edges. Edges describe which nodes are connected and nodes can have attributes which store specific information. In this context a graph representation of a molecule translates the connectivity of bonds to the edges of the graph. Molecular characteristics of the atoms (e.g., their name or residue name) are stored as node attributes. Before we detail the algorithm, it is handy to define a few more terms: We define a block to be a graph, which corresponds to all interactions and atoms of a single residue; Complementarily, a link describes the interactions (e.g., bonds or angles) introduced when two residues are connected. To this end, polyply internally uses the Networkx[41] and vermouth[42] python libraries for handling graph related computations.

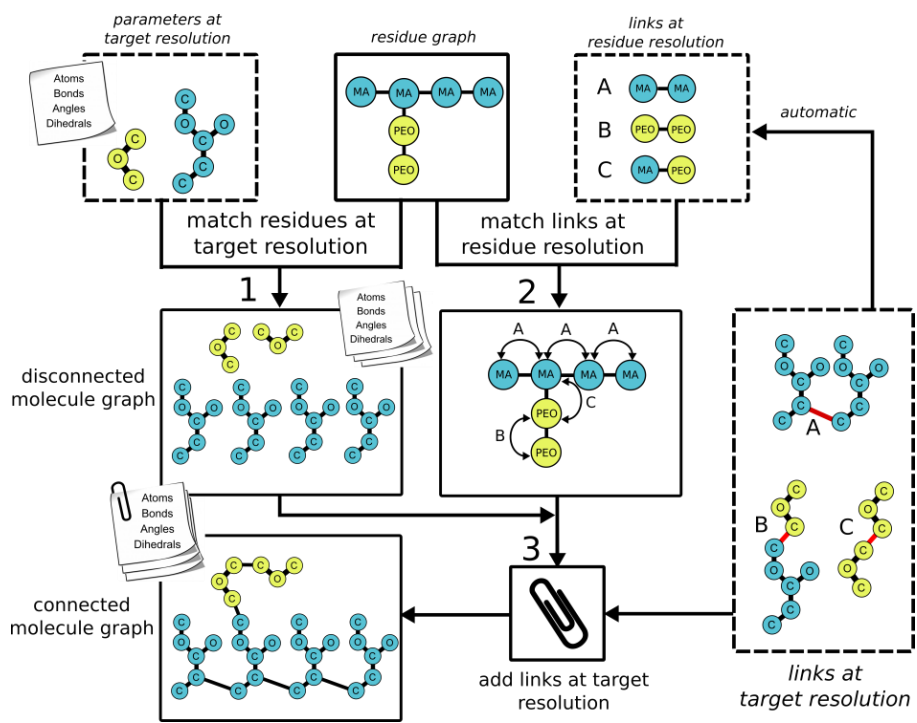

Figure 1. Schematic illustration of the workflow behind the parameter file generation exemplified for polyethylene oxide (PEO) grafted methyl acrylate (MA). The user input is the residue graph, while the building blocks are taken from the library (presented in dashed boxes). However, users can modify the library as they need. In step one the parameters for the blocks are applied based on the residue graph to form a disconnected graph at target resolution. In step two the links are matched at residue graph level to the input residue graph. Subsequently links are applied to the disconnected graph at target resolution in the third step producing the complete parameter file.

The general input to polyply for parameter file generation is a residue graph of the target molecule. In addition, the blocks and links corresponding to the residues in the target molecule are needed (cf. Fig. 1, dashed boxes). Currently, polyply is shipped with libraries containing these parameters for some force-fields and polymers, a database that will be expanded over time. From the definition of the residue graph, blocks, and links polyply generates a parameter file in three steps (cf. Fig. 1):

Step 1: Generate a disconnected graph of residues

After reading the input files, polyply iterates over all residues of the input residue graph. For each residue, the matching block is added to an empty graph thereby generating a disconnected graph of residues at target resolution. This graph already contains all atoms of the target molecule and interactions within the residues. This leaves as problem to assign the proper interactions linking the two or more residues.

Step 2: Find all links at residue level

To generate all interactions spanning more than one residue, links are applied between two or more residues. To solve this in a general manner we treat it as a subgraph isomorphism problem at residue graph level: we find all the ways a link can fit onto the residue graph subject to constraints such as matching node attributes. Performing this on the residue graph drastically reduces the problem size compared to solving the subgraph isomorphism problem at the target resolution. This establishes at residue level which links apply between which residues.

Step 3: Matching generic links to specific residues

Taking the matches between links and residues, the program establishes a correspondence between the atoms of the link and the atoms in the disconnected graph at target resolution. To do so the atom names and relative residue index given in the link are simply matched to the atoms of the residues in the disconnected graph generated in step 1. However, this matching step is not limited to the atom name and residue index. It can also be extended to take other atom characteristics into account. This allows to account for information that is not encoded in the connectivity of the residue graph, such as chirality or anomers of the same residue. When a link is added, also the edges of the link are added to the disconnected residue graph. In this way the disconnected graph gradually becomes a connected graph at target resolution level. This completes the graph transformation and the molecule including all interactions only needs to be written to a file.

## 2.2 – System building

Starting coordinates for systems are build using a generic multiscale approach, in which first a super CG resolution representation of the system is generated, followed by a back transformation to the target level. This multiscale approach is similar to the procedure underlying the Charmm-GUI polymer builder[30], but our approach is generic meaning parameters of the super CG model are derived on the fly based on the target force-field, employs a self-excluding random-walk in contrast to a full-scale dynamics simulation, and uses an automated back transformation, which does not rely on library of coordinate fragments.

The system building proceeds in five steps (cf. Fig. 2):

Step1: Mapping all molecules to one-bead per residue

In the first step, the topology file is analyzed and all molecule types in the system are detected. For each molecule all unique residues are identified and converted to blocks. A generic one bead per residue super CG model is created and stored in the form of a graph. The underlying connectivity of the residue graph is extracted from the bonded graph of the molecules.

Step 2: Generate coordinates for residues

Each block is a graph of a single residue and graph embedding is used to generate coordinates for this residue. Due to the specific requirements of molecular geometry, we utilize a two-step graph embedding. First, initial coordinates are generated using the Kamada-Kawai[43] embedding as implemented in the NetworkX library[41]. Subsequently, we perform a geometry optimization based on the bonded interactions within the residue using the scipy[44] implementation of the Limited-memory Broyden–Fletcher–Goldfarb–Shanno minimizer[45].

Step 3: Derive parameters for the generic CG model

In the self-excluding random walk a one-bead per residue approximate CG model is used. It is based upon a Lennard-Jones (LJ) potential as the interaction function. The epsilon parameter of the model (LJ well depth) is always set to 1 kJ/mol. Since we do not perform dynamics the attractive part of the potential is less important. The sigma parameter, however, determines the overall packing density and is computed from the residue template coordinates, reflecting the volume of the residue. It is derived from the radius of gyration as detailed in the supporting information. In general, the radius of gyration is often used in polymer physics to estimate the spherical volume occupied by a single chain. Here we port this concept to molecular geometry of a single residue. However, in addition to the geometry of the residue we also account for the fact that the single atoms have a certain volume. Although it is an approximate radius and probably not useful in an actual simulation, we find it to be good enough in the context of the random walk.

Step 4: Constrained random walk

To generate coordinates for the one-bead per residue molecules in our target system, we perform a self-excluding random walk. An attempt to place a bead (step) in the random walk is rejected, if a maximum force on the placed bead is exceeded. The self-excluding random walk is performed along a breadth-first traversal of the molecular graph. This means nodes (i.e. residues) that are close to each other are placed first and then the algorithm proceeds further along the chain. Molecules are placed separately after each other, where the starting point is randomly chosen from a grid.

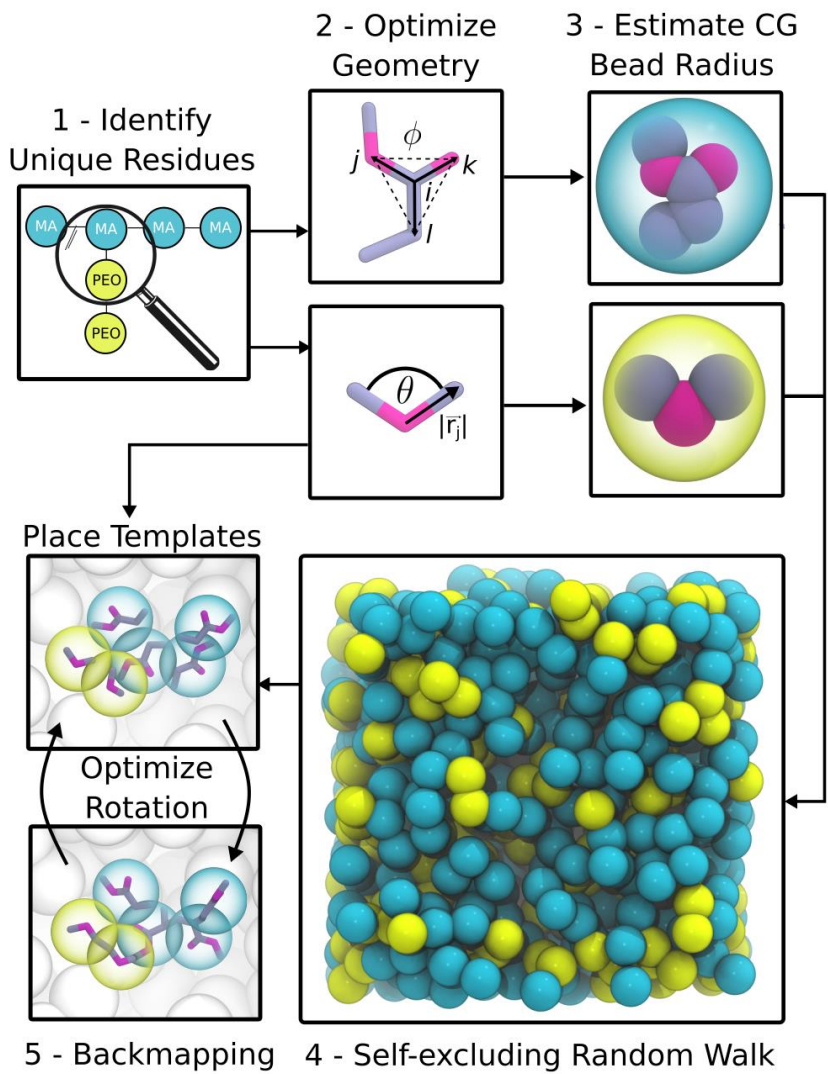

Figure 2. Principle of the multiscale algorithm for building systems. The algorithm works in five steps: first residues are identified, followed by generating coordinates for the residues through two-step graph embedding. Based on the residue volumes the generic CG model is generated, that is then used in a self-excluding random-walk. Finally using the template residue coordinates the CG model is backmapped to the target coordinates.

This grid can either be user-specified or is considered to be rectangular across the box. When a the random-walk algorithm exceeds a certain number of steps, it goes backwards in the breadth-first path by default by 10 residues and tries to replace these 10 residues. In addition to the force-criterion acceptance, the random-walk attempt can also take into account geometrical constraints (i.e., regions of space that are excluded from sampling). Furthermore, it is possible to limit the dimensions of the random walk to specific directions to steer it and generate more straight chains. This is especially useful for generation of polymer brushes as the chain conformations are more extended compared to a melt. All interactions are computed considering rectangular periodic boundary conditions using the scipy c-implementation of the KD-tree.[44] Using a KD-tree makes it possible to compute a large number of distances within a cut-off efficiently within python.

Step 5: Backmapping

Low resolution coordinates are transformed to the higher resolution target coordinates by a residue template based backmapping procedure similar to those used for biomolecules[46]. First, the center of geometry of the residue template is moved to the CG position. Subsequently we optimize the rotation of the template around the center of geometry such that atoms which have a bond to other residues are placed close to those residues. To do this we first perform a connection analysis finding which atoms connect to atoms of the neighboring residues. If the target resolution coordinates are available, they are used over CG coordinates. Overall, this aligns the residue template with the chain back-bone in a generic fashion. This procedure is also applicable to branched residues and optimizing the rotation is efficiently done in python. Similar to the idea in the backwards program[47] the template residue coordinates are scaled by a fudge factor of 0.45 from the center of geometry. In the final step, the user has to run a regular geometry optimization. This causes the coordinates to relax to a final state that can be used as input to run an equilibration simulation. To implement chirality we currently use a specific improper dihedral that forces chirality during the energy minimization step.

## 3 – Example Applications

In this section we will demonstrate the capability of the program with some examples for realistic systems. In general, we will briefly present the steps required to generate both input parameter files and coordinates. For each system, we ran a small simulation to confirm the parameters are correct and the system is stable.

### 3.1 – Polymer melts at atomistic and CG level

Simulation of amorphous polymers or melts are the backbone of much research in material science and polymer science in general. Depending on the length and time scales that are needed atomistic resolution or CG models are utilized. Here we demonstrate that polyply is capable of generating realistic melt conformations independent of the target force-field, and we analyze it's performance. In particular, we have generated melt systems of varying sizes for poly methyl acrylate (PMA), polystyrene (PS), poly methyl methacrylate (PMMA), poly vinyl alcohol (PVA), polyethylene (PE), and PEO. For each system we generated the required parameter files and initial coordinates for 100 chains of length 50, 100, 250, and 500. Each of these systems were generated at atomistic level using GROMOS and CG level using the Martini3 force-field. Figure 3A shows the atomistic structures of the six polymer species surveyed and overlaid as circles those atoms treated as one particle in the CG models. After system generation with polyply, we ran an energy minimization and computed the average end-to-end distance of the polymers from the minimized configuration. To improve statistics, 10 replicas of each system were generated leading to a total of 480 systems. The initial target densities are reported in the table S2.

In a melt, polymer conformations can be described by ideal chain statistics. Some properties like the end-to-end distance can be computed from theory utilizing experimental input quantities such as the characteristic ratio. To show that the initial polyply conformations are realistic for melts, we computed the expected end-to-end distance using two models – the hindered rotation model (HRM) and the worm-like-chain model (WCM) – and compare those distances to the end-to-end distance generated by polyply. Figure 3B shows the comparison. Overall, the polyply structures clearly follow the trends of both models, but also the quantitative agreement is good with mean absolute errors of about 5.5Å compared to the HRM and 3.0Å compared to the WCM. For the WCM we also compared the distributions obtained from polyply and the model as shown in figure S1. We find a good qualitative agreement.

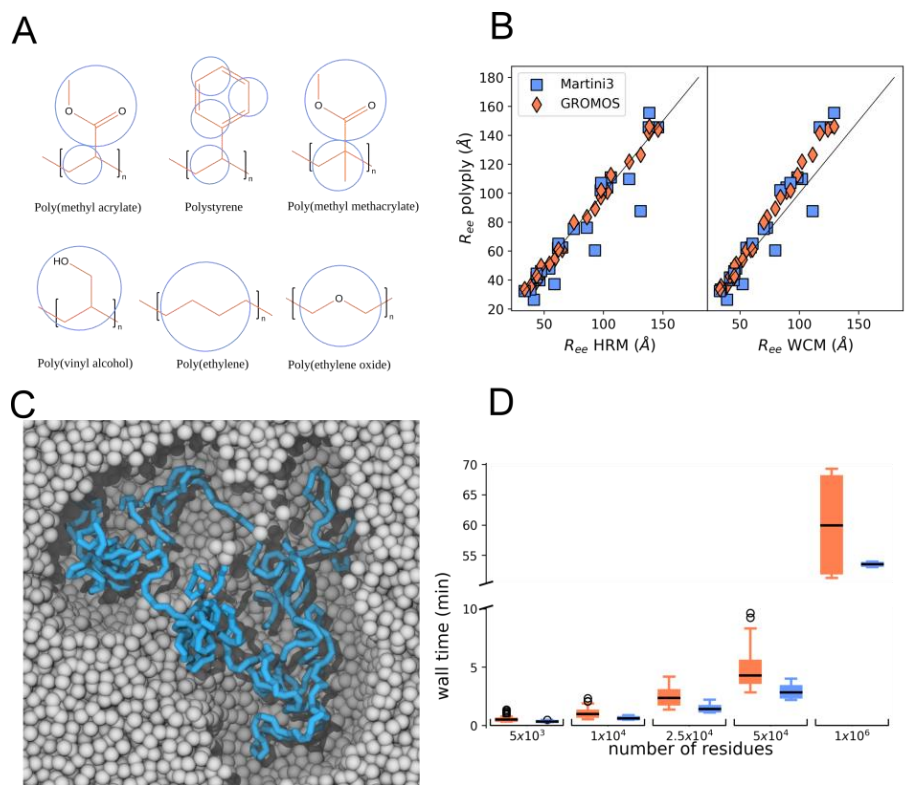

Figure 3. Characteristics of melt systems generated by polyply and performance. A) Atomistic structure and mappings to Martini level for all six polymer species surveyed. B) End-to-end distance of the melt structures generated with polyply compared to those obtained by theory using the HRM (left) and WCM (right). Blue squares indicate Martini structures, and orange diamonds GROMOS. For each of the polymers, systems with four different chain lengths (50, 100, 250, 500) were built. C) Single melt chain of Martini3 PEO with 500 residues (blue), other chains are shown in gray and residues within 1 nm are omitted for clarity. D) Typical time for generating coordinates with polyply for different total numbers of residues. The performance is shown per force-field (blue: Martini3, orange: GROMOS), averaged over all systems shown in panel B. In addition, systems with 1 million residues have also been set up as further benchmark. Atomistic performance outliers are all from melts of PE.

It should be noted that of course the density and end-to-end distance of the force-field can be different from experiment or theory. The end-to-end distance can be further fine-tuned by optimizing the step length of the random walk. However, even if not optimized, relaxation to target density and end-to-end distance is typically observed within 50 ns or less, as shown in figure S2. for a subset of systems. Figure 3C shows a single chain of PEO 500 in a melt after 5 ns of simulation. Other chains are removed within a 1 nm radius around that chain to illustrate the coil like conformation of the chain in the melt. Figure 3D shows the typical time it takes to generate coordinates for different number of total residues using polyply. Generating CG residues is slightly faster than generating atomistic residues. Overall melts of up to 50,000 residues are typically generated in less than 5 minutes. Larger systems in the order of 1 million residues are generated within less than 90 minutes.

## 3.2 – Polymeric lithium ion battery

Ion conducting polymers for the application in lithium ion batteries have been a very active field of research for many years.[11,35,48] Block-copolymer systems, consisting of one conducting polymer and one polymer improving the mechanical stability, are a very promising route to new and enhanced batteries. Simulation of ion conduction in these systems, however, are less common as creating the initial coordinates poses serval technical challenges: 1) The system has to be obtained in the phase separated state. Modeling salts in common bead-spring models or mean-field theories is difficult when the effect on phase separation is unknown. 2) The salt has to be distributed within the PEO layer but without generating overlaps. Especially, for the commonly used large anions this becomes problematic. 3) In principle, the material is best represented by a multilamellar systems, in which for example cross conduction can also be observed even though for PS-b-PEO this is known to be less problematic. In this example we will show how to create a multi-lamellar system of PS-b-PEO doped with lithium bistriflimide (LiTFSI) using realistic polymer length as well as representing the ions explicitly. Our target system comes from the experimental work of the group of Balsara, who have exhaustively studied PS-b-PEO doped with LiTFSI.[49]

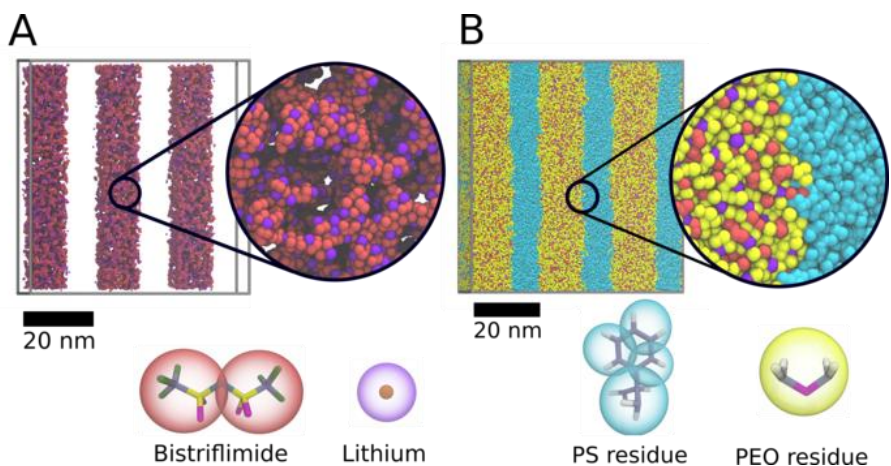

Figure 4. Phase separated block-copolymer PS-b-PEO doped with Lithium Bistriflimide (Li-TFSI) generated with polyply. A) Li-TFSI salt placed within the domains where PEO is going to be located. The zoom provides a more detailed view on the salt also showing that there is empty space in the initial dispersion. B) Structure obtained after growing in the block-copolymer around the salt shown in panel A and running 50ns of equilibration. The generated morphology contains six lamellar regions. The zoom shows an in-depth view onto the interface between the two polymers, showing that empty space has been filled. Below both panels, the mapping of the all-atom residues to the CG level are shown.

The block-copolymer in this example consists of PS with a molecular weight of 6.4 kg/mol (~63 monomers) and PEO with a molecular weight of 7.3 kg/mol PEO (~163 monomers). LiTFSI is mixed in using a ratio of Li to PEO monomers of 0.085.

From the experimental work it is known that this composition forms a lamellar phase with a domain spacing of 20 nm. The Martini input files for the polymers are simply generated using polyply gen_params and our Martini library of polymers. To generate the starting structure in the phase separated state, we specify rectangular geometric restrictions on where each part of the block-copolymer is allowed in the box. To that end, we define six alternating domains of 12 nm and 6 nm sizes in which PEO and LiTFSI or PS are allowed, respectively. The domains are unequal because the volume fractions of PS and PEO are not the same. To provide the Martini consistent domain spacing, we generated a system with a single lamella and equilibrated the volume. The thin film is generated in two steps, aiming at a uniform salt distribution in the PEO part. In a first step, the salt is dispersed throughout the box placing it inside the domains where PEO is allowed. Figure 5A shows the salt as well as the domain boundaries as obtained after this step. Subsequently, we generate a grid of starting points on the boundary of these domains. Starting on these grid points the chains are grown into the domains by our random walk using the generic super CG model. This approach has previously shown to be adequate for simple bead spring models as well.[34,50] Finally the program backmaps the structure to Martini target resolution. Once the starting structure is generated, which typically for this size of systems takes 30 minutes, an energy minimization is performed. Subsequently a short equilibration of 5ns keeping the z-dimension fixed and only applying pressure coupling in xy is performed to allow for chain packing to increase orthogonal to the stack direction. Subsequently we ran a 50 ns equilibration under constant area only coupling the z-direction to equilibrate the salt distribution. Figure 5B shows the morphology of the system after this short equilibration phase. The thin film has a size of about 60 nm x 60 nm x 10 nm and comprises roughly 600,000 particles. As shown by the zoomed view as well as the density profiles (SI), the space is completely filled with the polymer and salt.

## 3.3 – Lipid vesicle with liquid-liquid phase separated interior

Liquid-liquid phase separation (LLPS) is an important driving force in both biotechnological applications and biological systems. Systems capable of undergoing LLPS are therefore of high interest to many researchers and are not only studied experimentally but also at various levels of theory.[10,51] Concerning cellular processes, LLPS is speculated to have promoted the early stages of life by allowing to form simple compartmentalization, which eventually lead to the evolution of membrane-less organelles inside modern day cells.[52,53] As such, studying LLPS in the context of cellular environments is of considerable interest. While the supporting programs for bio-molecular simulations allow to generate cell membrane structures of entire mitochondria or virus envelopes, filling those with anything else than water and ions is usually challenging, especially when polymer phases are to be simulated. In this example, we setup a system consisting of a multicomponent lipid vesicle, composed of dioleoyl-phosphatidylcholine (DOPC), dipalmitoyl-PC (DPPC), and cholesterol, and containing PEGylated 1-palmitoyl-2-oleoyl--phosphatidylethanolamine (POPE) lipids, filled with a phase separated aqueous solution consisting of PEO and dextran in the interior. This system has been experimentally shown to induce vesicle fission and therefore gives insight into the generation of early life [54]. Similar systems are also considered important to the development of synthetic cells[55].

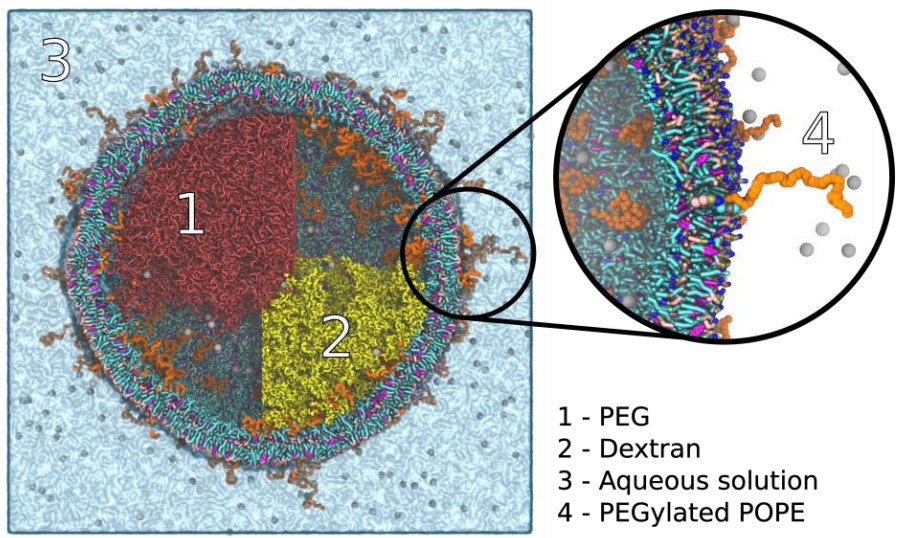

Figure 5. Sliced view of a lipid vesicle composed of DOPC, DPPC, Cholesterol, and PEGylated POPE (PEG part in orange), containing a coacervate of PEO (red) and dextran (yellow). Inside and outside the vesicle is water and sodium to counterbalance the negatively charged PEGylated POPE. The diameter of the vesicle is 40 nm and the left half is filled with PEO and the right with dextran. Some polymers in the upper left and lower right are omitted to show the ions and PEG tails of the PEGylated lipids on the inside. The zoom shows more detail of the membrane showing a PEGylated lipid extending from the bilayer surface.

This example not only demonstrates that polyply interfaces well with the area of biomolecular simulations, but also demonstrates several technical challenges: 1) PEGylated lipids require the addition of PEG to the lipids of the vesicle and need to be placed without penetrating the bilayer; 2) dextran is a branched sugar polymer, which typically has a statistical distribution of molecular weights and branches; 3) the coacervate needs to be generated in the phase separated state. To achieve this challenge, first, parameter files for PEGylated lipids and PEO are generated using the Martini library and the gen_params tool. The molecular weights of PEO in the vesicle and in the PEGylated lipids are 2000 g/mol (~45 monomers) and 8000 g/mol (~180 monomers), respectively. Generating parameter files for dextran is more complicated, however.

Dextran is a polysaccharide composed of α1,6 linked glucose residues with α1,3 connected branches. In addition, it is in general polydisperse and the branching depends on the molecular weight. For the target molecular weight of 10000 g/mol, dextran has on average 5% branches from the main chain with a length of up to three residues.[56,57] To model the diversity in dextran's molecular structure we generated 500 residue graphs, with random number of branches and lengths as outlined in the Supporting Information. Using this distribution of structures (Fig. S4) polyply gen_params was used to create parameter files for all those structures, which took less than 60 seconds.

To generate starting coordinates for this system, we first obtained a vesicle using TS2CG[24]. As there are more specialized programs to generate lipid bilayers in various shapes and forms it was not our intention to also generate those using polyply. The lipid coordinates generated by TS2CG were given as starting structure to polyply. In addition, a geometric constraint was used to specify that PEO and dextran can only occupy half of the vesicle, approximating it as a sphere, with a region of 2 nm overlap to allow some interphase mixing. With this input, the system is generated by our generic super CG random walk followed by a backmapping step. Polyply also automatically identifies that PEGylated lipids have to be extended, as the random walk algorithm will steer the conformations away from the membrane. Generating the entire system took about 30 min on a single CPU. Once the initial coordinates are obtained, an energy minimization was run, followed by a short equilibration and 50 ns of production run. The system was stable, and its final configuration is shown in Figure 6. We clearly see that the system remains phase separated at least over the timescale simulated and that the PEGylated lipids are uniformly distributed both on the inside and outside of the bilayer.

## 4.0 – Discussion

The increase in computing power propels MD simulations of systems into new areas, which in their complexity or size were unmanageable only few years ago. However, equilibration of such systems far from the desired equilibrium state takes time and is costly in terms of computing power. A second complicating factor relates to generating input files for such simulations, as setting up the input parameters not only takes human time but is also intrinsically error prone. The latter aspect becomes especially problematic when HT approaches are being designed. Programs which facilitate HT MD simulations featuring multicomponent complex systems are therefore highly desirable. While those exists for biomolecular simulations, they are largely missing in the field of material science. To resolve this situation, we have defined five major challenges a program needs to be able to solve: 1) it needs to be resolution and force-field independent; 2) able to setup complex and large systems; 3) support arbitrary complex molecular topologies (i.e., branching); 4) be able to generate complex morphologies; 5) and finally be reasonably fast to permit HT research.

In this paper, we have presented polyply, a software suit that is aimed at facilitating simulations involving polymers at any target resolution and force-field desired. Parameter file generation and coordinate generation are split into two independent programs – polyply gen_params and polyply gen_coords. The gen_params program implements the graph transformation used within polyply to generate parameter files. The algorithm takes a residue graph and maps it into a higher resolution graph, which is agnostic to the target resolution thereby meeting the first of the challenges outlined above. It further matches fragments, which describe the bonded interactions between residues, by finding all subgraph isomorphisms between these fragments and the residue graph provided. Therefore, it will also assign parameters for complex branched polymers This has been shown to work based on our test cases. Especially the last test case exemplified this aspect, where polyply was used to generate a statistical distribution of the branched dextran polymer that involves two different linkage types between otherwise equivalent residues. As outlined in the challenges, polyply is also compatible with input from biomolecular simulations, which we have shown previously by using polyply as part of a protocol for setting up PEGylated proteins.[58] In addition to supporting large complex polymers, generating input parameter files is also fast, making it suitable for HT applications. For example, generating a parameter file for an atomistic polystyrene chain of 1000 residues, which involves more than 100,000 bonded interactions, takes less than 10 seconds. Overall, we showed that the polymer parameter generation meets all requirements outlined in the challenges. Currently, the main limitation of the parameter generation is that it is limited to GROMACS input files. However, extension to other MD software is possible and would only require additional input file-parsers as the core of the code makes no calls to the GROMACS software itself.

As further outlined in the challenges, structure generation is a key step for any program. It needs to be fast and generate complex structures that are good enough to start the simulation without extensive relaxation. To this end, polyply gen_coords implements a multiscale approach to generation of condensed phase systems at near to target density. The multiscale approach is based upon using a generic one-bead per residue CG model, performing a self-excluding random walk, and backmapping it to target resolution. The interaction potential used for the self-excluding random walk, which essentially approximates the volume of a residue, is directly computed from the geometry of the residue as well as the force-field parameters of the target polymer. Thus, our multiscale approach can be used independent of the force-field and target resolution. To further validate the robustness and quality of this approach, we generated 480 melt systems at two levels of resolution for four polymer species. For all systems an energy minimization could be run without failure. Not only were the systems stable, the average end-to-end distance of the initial frames averaged over 10 replicas compared very well to theoretical calculations for melts. This suggest that the distribution of conformations and entanglement produced by polyply are realistic, providing a good starting point close to the equilibrium configuration. This is further confirmed by running test simulations for some replicas, which show small relaxation of the end-to-end distance to the force-field specific value as well as to the force-field specific target density.

Using a one-bead per residue CG model not only makes the approach force-field and resolution independent, but it also greatly increases the performance. We assessed the performance for 360 melt systems up to 50,000 residues. For polystyrene at Martini level 50,000 residues equates to 200,000 coordinates and at GROMOS level to 600,000 coordinates. These systems are typically generated in less than 5 minutes independent of the chain length and target force-field. This performance compares favorably to the recently published PyPolyBuilder, which takes about 8 minutes for a single chain of 1248 coordinates.[29] To further benchmark the performance of the structure generation, we also set up systems with 1 million residues, corresponding to 13 million coordinates for atomistic polystyrene. These are generated on average within less than an hour on a single core.

As the generic multiscale approach is very efficient in packing even long polymer chains it was possible to augment it with further criteria used in the random walk chain placement. For example, it is possible to force the direction of the random walk, which makes it very suitable for generation brushes. An example is outlined as part of our online tutorials. Furthermore, the multiscale coordinate generation uses a random walk breadth-first traversal of the molecular graph. This means residues which are connected by bonded interactions are placed close in space, making it possible to generate even complex branched structures. In addition, simple geometrical constraints can be utilized to build phase separated systems. By generating a liquid-liquid phase separated system consisting of highly branched dextran and PEO inside a vesicle as well as a micro phase separated block copolymer PS-b-PEO system, we have shown that these tools can generate a variety of inhomogeneous systems. The former example also shows that it can easily be combined with already existing systems, especially those for which more specialized builders exits, such as lipid membranes. Overall, these examples demonstrate that polyply is capable of setting up large and complex systems, with an excellent performance in a realistic but force-field and resolution independent manner.

However, the coordinate generation also has some limitations. Generating macro-cycles, for instance, is currently not possible. Although it is interesting to study macro-cyclic polymers, for example some viral DNA, it is no possible to generate input structures within the scope of a self-excluding random-walk. A further limitation is that currently the structure generation only supports rectangular PBC conditions. Although they are sufficient for a wide variety of systems, more complex PBC conditions could be implemented by extending the scipy cKD tree, which is the workhorse for interaction computations. Finally, while simple geometric restrictions for the random-walk work fine for many applications, extending them to arbitrarily shaped boundary surfaces would enable to directly read in experimental density maps and then grow polymers on top of those. This would require some form of triangulated surfaces to be used as boundary surfaces, as is done in TS2CG[24].

In conclusion, we have demonstrated that the software suite polypy, presented here, is able to generate input files and starting coordinates for complex and challenging systems, connecting the bio-molecular world to material science.

## 5.0 – Methods and Models

**MD settings.** All MD simulations were done with GROMACS[59,60] (versions 2020 and 2019), using the Verlet cut-off scheme. For the CG simulations, a cut-off of 1.1 nm was used for the LJ interactions, whereas the Coulomb interactions were computed within a cut-off of 1.1 nm, and PME for longer-range contribution to the electrostatics whenever the system at hand contained charges. The energy minimizations were done with the standard steepest descent algorithm as implemented in GROMACS. The CG MD simulations were performed using the default leap-frog integrator with a time-step of 20 fs in the isobaric-isochoric ensemble. Temperature was kept constant using the v-rescale algorithm[61] and the pressure was coupled using the Berendsen barostat[62], typically used for equilibration of MD simulations.

The atomistic MD simulations were conducted with a cut-off of 1.4 nm for the Lennard-Jones interactions and electrostatic interactions. Long range electrostatics were further treated with the reaction field method, where the dielectric constant was set to 2 for all polymer systems, which is reasonable considering the typical low dielectric constant of vinyl polymers. Energy minimizations were conducted as well with the steepest descent algorithm and MD simulations were run with the default leap-frog integrator using a time-step of 1 fs for the integration. The time-step of 1fs is necessary as the relaxation runs to target density are done using unconstrained bonds. For production runs the time-step can be increased to 2 fs with constrained bonds, in principle. As explained for the CG simulations also the GROMOS simulations were run in the isobaric-isochoric ensemble using v-rescale[61] temperature coupling and Berendsen pressure coupling[63].

**Systems.** Polyply was run for all test-systems within python3 (v3.6.9 on local machines, and v3.8.2 on the Dutch National Supercomputer, Cartesius). For acceleration the numba package[64] was installed in all environments. All test systems were either run on a local desktop machine running Linux OS or on the national HPC cluster. The performance for parameter file generation and input coordinate generation were recorded on the HPC cluster for the CG systems, running 24 processes in parallel on a single node. On the other hand, atomistic benchmark times were recorded on a desktop machine running 10 processes in parallel.

**Models.** The all-atom polymer models were implemented following the rules for creating polymers within the GROMOS 2016H66 force-field and charges adopted from similar functional groups as is custom within GROMOS.[38] The current library does not only support homopolymers but all combinations of monomers. Martini CG models were parametrized newly following the Martini 3 guidelines for making molecules or were adopted from existing Martini 2 models. Each new model was based on the GROMOS model, for which a system was generated with poyply and run with the settings stated above. We note that these models are subject to future improvement and that currently only homo-polymers are supported with the exception of PS and PEO block-copolymers. A detailed validation of these models will be provided in a separate publication. The current (beta) versions are available from the polyply library see code availability.

## 6.0 – Acknowledgments

We would like to thank the Center for Information Technology of the University of Groningen for their support and for providing access to the Peregrine high performance computing cluster. S.J.M acknowledges funding from the ERC via an Advanced grant "COMP-MICR-CROW-MEM", grant agreement ID 669723. Computational resources for this work were partly provided by the Dutch National Supercomputing Facilities through NWO.

The authors thank:

M. C. Ramos, Y. M. H. Gonçalves, and B. A. C. Horta, for checking the GROMOS input parameters and discussions on modeling polymers with the 2016H66 GROMOS force-field.

T. A. Mayer and A. J. Gormley for discussions on high throughput polymer synthesis and the future potential of simulations to enhance experimental high throughput protocols.

Melanie König for her feedback and perspective on the figure design and the writing.

## 7.0 – Code and data availability

Polyply is distributed via the pypi package index (https://pypi.org/project/polyply/) and is developed publically on GitHub (https://github.com/marrink-lab/polyply_1.0) under the permissive Apache-2.0 license. The polymer model parameters and input files used in the examples are part of the polyply library and/or tutorial repository included with the code.

## 8.0 – References

# Supporting Information

Fabian Grünewald[1], Riccardo Alessandri[1,2], Peter C. Kroon[1], Paulo C.T. Souza[3], Siewert J. Marrink[1]

[1] Groningen Biomolecular Sciences and Biotechnology Institute and Zernike Institute for Advanced Materials, University of Groningen, Groningen, The Netherlands

[2] Pritzker School of Molecular Engineering, University of Chicago, Chicago, IL 60637, USA

[3] Molecular Microbiology and Structural Biochemistry, UMR 5086 CNRS and University of Lyon, Lyon, France

**Contents**

**Polyply polymer library**

*Table S1. Overview of all polymers implemented in the polyply polymer library with corresponding abbreviations and allowed combinations. All combinations means all monomers can be combined with all other monomers in the library, homo refers to a polymer only being implemented as homopolymers.*

| Name | Abbreviation | GRMOS combinations | Martini3 combinations |
|---|---|---|---|
| Polyethylene glycol | PEO | all | PS, lipids |
| Polystyrene | PS | all | PEO |
| Polymethyl acrylate | PMA | all | homo |
| Polymethyl methacrylate | PMMA | all | homo |
| Polyethylene | PE | all | homo |
| Poly(3-hexylthiophene) | P3HT | all | homo |

| Polyvinyl alcohol | PVA | all | homo |
| Dextran | DEX | not included | homo |

## Sequence generation auxiliary tool

Polyply generates polymers from a residue graph. For simple linear polymer that is simply the sequence , however, branched polymers or polymers, which have a statistical distribution cannot simply be described by providing a linear sequence. Whereas for regular linear polymers polyply allows to provide the sequence directly as command line argument (-seq) for the gen_params tool, more complex residue graphs can be provided in the form of a networkx graph json-file. To further facilitate generating parameters for these more complex polymer topologies we have implemented the 'polyply gen_seq' auxiliary tool. The gen_seq tool can be used to create and manipulate residue graphs from the command line interface. Here we briefly describe the general usage and logic behind the program.

The command line input to gen_seq are strings, which define macros. Macros are graphs and represent fragments of residue graphs. Multiple macros can then be combined to form the residue graph of interest. For example, one macro can describe a linear fragment and another one a branched fragment. Combining them gives a residue graph describing a Janus like particle. Macros are defined as strings, with the following syntax:

<macro_name>:<#blocks>:<#branches>:<residues>

The <Macro_name> field is an arbitrary name of that macro, the <#blocks> field should be the number of residues for liner graphs and the number of generations for branched polymers. Consequently, <#branches> is the degree of branching of that polymer. Only regular branching is supported. The <residues> field can be used to define what the residue composition is within that macro using the following syntax:

<residue_name-probability, other_residue_name-porbability>

The residue name can be any string and the probability is the probability of having that residue anywhere in the chain that is described by the macro. Let us consider some examples for making macros. To generate a linear chain of 10 monomers PEO one would use 'A:10:1:PEO-1.0'. Note that now the macro is called A. If we want instead to have random distribution of PS and PEO with equal probabilities we would use 'A:10:1:PEO-0.5,PS-0.5'. In that case PS and PEO are randomly distribution along the graph with equal probability. In contrast if we want to generate a polymer that is tree like and consists of poly(propylene imine) (PPI) residues, we would use 'A:4:2:PPI-1.0'.

Once the macros are defined the can be combined to form a residue graph. To do so first the macros that are to be used have to be enumerated using the –seq flag. This specifies the order of the macros in the final graph. Subsequently, the connections between the macros have to be defined using the '-connects' flag and following syntax:

<seq_position_macro_A>,<seq_position_macro_B>-<connect_residueA>,<connect_residueB>

In this syntax <seq_position> is the position index of the macro provided to the –seq flag, and <connect_residue> is the residue ID of the residue that form the connection between macros A and B respectively. In addition to these basic functionalities the program also allows to attach tags to each residue in the sequence with the -label option. In this way, for example, nodes can be labeled with their chirality/tacticity. There is also the possibility to modify end-groups. Both options are described in more detail in the help menu of the program. As polyply parses networkx style json-file. Thus users can write their own protocols for making residue graphs or edit them manually as well.

**Estimating the volume from the radius of gyration**

In polyply, the volume is computed directly from the radius of gyration of the residue, however, taking into account the van-der-Waals radius of the atom/bead. The formula for the radius of the volume becomes:

$$r_{res} = \sqrt{\frac{1}{N}\Sigma_k(1 + r_{vwd}) * (r_k - r_{COG})} \quad (1)$$

Essentially we compute the radius of gyration but add a van-der-Waals contribution in the direction of the position vector from the center-of-geometry (COG). This makes sure that we capture some volumetric component in the radius of gyration. This procedure was selected for two reasons: 1) the radius of gyration is essentially a moment of inertia neglecting the mass. Thus it captures a radius that the residue has when rotating about the center of geometry. In other words it could be described as a rotational volume as opposed to the real volume of the residue. When generating an initial starting structure we need to make sure that there is no direct overlap between any atoms when transforming from the super CG model to target resolution. However, using the actual volume as for example computed from a voxelized approach that sums the volume of the intersecting Vdw-spheres, we frequently find overlaps. 2) The above description can take into account larger ligands, conjugated systems, and residues which are much larger than a typical polymer radius. Imagine computing the volume for C60, which is completely round but hollow inside. If we were to compute the volume using a voxelized approach by summing the Vdw radii we would greatly underestimate the volume. The same holds for example for a lipid that is treated as one residue or an extended conjugated system. Ultimately it is a practical choice where we observe that our radius of gyration approach works well.

**Computing end-to-end distances from theory**

End-to-end distances were computed form two models the worm-like chain model (WCM) and the hindered rotation model (HRM). In case of the WCM model the distribution was computed as suggested previously by Lee et al. and subsequently the expectation value of the distribution was calculated to obtain the average end-to-end distance.[1] In the case of the HRM the mean square end-to-end distance is given by equation 2.[2]

$$<R^2> = n * l^2 * C_\infty \quad (2)$$

In equation 2 $C_\infty$ is Flory's characteristic ratio, which can be measured experimentally or computed from the persistence length. Furthermore, $n$ is the number of bonds with length $l$. The distribution of the WCM model is computed from the persistence length ($l_p$), which was obtained from Flory's characteristic ratio (c.f. eq2) and the contour length, which was computed from the bond length ($l_b$) and angle ($\vartheta$) of the atomistic model parameters (c.f. eq3 and eq4).

$$lp = \frac{C_\infty}{2*l} \quad (3)$$

$$R_{max} = n \times l \times cos\frac{\theta}{2} \quad (4)$$

Sometimes, for the same polymers several different values for the persistence lengths and characteristic ratios are reported in literature. Whereas for polymers such as PS the values of the characteristic ratio range from 8.24Å[3]-9.6Å[4], which is rather close together, for others such as PEG or PVA quite different values are reported. For PEG the values range between 4.2Å and 11.2Å.[5] However, there appears to be an agreement between simulations and theory[3,5–7] for a lower value between 2Å-6Å. These values fit better with the persistence lengths and end-to-end distances obtained from simulation and theoretical calculations. Therefore, we have taken the average of three theoretical calculations[5] as reference value. Similarly for PVA estimates range from 3.4Å-8.5Å[3,8]. In this case it is unclear which value is more reliable. We have chosen the value of 4.81Å as recently reported for CHARMM all-atom simulations[3], as that was recorded for the melt. Bond lengths and angles were taken from the GROMOS 2016H66 model, for which the carbon-carbon bond length is 1.53Å and the angle theta is 70 degrees. For PEG the average of the C-O and C-C bond was taken as is custom. Table S2 collects the values for the characteristic ratio, target density together with the corresponding temperature as well as bong length and theta angles.

*Table S2. Physical constants of polymers used in the end-to-end distance calculation. Densities at 413K are taken from ref 2 and densities at room temperature are the amorphous densities taken from ref [9]. $C_\infty$ values are averaged over several references unless stated otherwise. *average of theoretical values reported in ref 5.*

| polymer | $C_\infty$ (Å) | T (K) | ρ (g cm$^{-3}$) | θ degree | $l_b$ (Å) |
|---------|----------------|-------|-----------------|----------|-----------|
| PS | 9.11[2,3,10] | 413 | 0.969 | 70 | 1.53 |
| PE | 7.34[2,3] | 413 | 0.784 | 70 | 1.53 |
| PMA | 8.15[3,11] | 298.15 | 1.100 | 70 | 1.53 |
| PEO | 4.5* | 413 | 1.064 | 70 | 1.48 |
| PMMA | 8.2[2] | 298.15 | 1.100 | 70 | 1.53 |
| PVA | 4.81[3] | 298.15 | 1.200 | 70 | 1.53 |

**Comparing end-to-end distance distributions from polyply to the WCM**

As outlined in the main text, we find a good match between the average end-to-end distances generated by polyply and the two models presented previously. For the WCM, however, we not only have information about the predicted average values but also the distribution associated with that model. Figure S1 compares the distributions obtained by the WCM model and polyply for PEO using the GROMOS force-field (panel A) and Martini force-field (panel B) for all four chain lengths studied. We have chosen to use PEO as an example, because it has been previously reported that PEO is well described by the WCM model.[12] Overall the agreement is good, but we note that the distributions of the AA model is slightly shifted for the longer chain lengths. This is a common trend we see, however, it should also be noted that in comparison to the WCM model the higher molecular weights are worse, whereas in comparison to the generally more accurate HRM model the average end-to-end distances are in better agreement.

**Density and end-to-end distances from MD**

For each of the GROMOS 50 repeat unit melt chain systems, we ran a 10ns MD simulation under constant pressure. Figure S2A) shows the densities and S2B) the end-to-end distances as function of simulation time. The temperature for all simulations was 513K to ensure that all simulations are well above the melting point of the polymer species. For each of the MARTINI 50 repeat unit systems we ran 50ns of MD simulations. The corresponding results are shown in panel C and D. As apparent form panel A, and C the end-to-end distance quickly relaxes to the force-field native average end-to-end distance and then fluctuates around the average. However, no strong relaxations can be observed suggesting the end-to-end distances from polyply are a good starting point. Similarly as shown in panel B, and D the densities take some ns to relax to their target values at the temperature of interest, which typically for this system size takes less than 1 nanosecond.

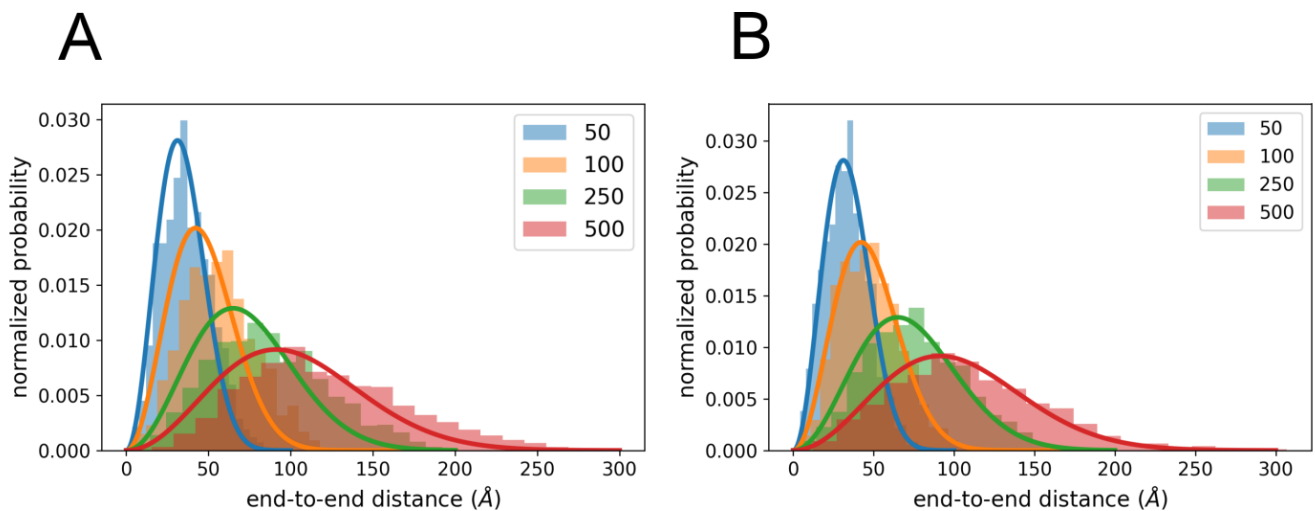

*Figure S1. End-to-end distance distribution of PEO obtained from the WCM (solid curves) and polyply (histogram) for 4 chain lengths. Panel A shows the GROMOS based distributions and panel B those obtained for Martini.*

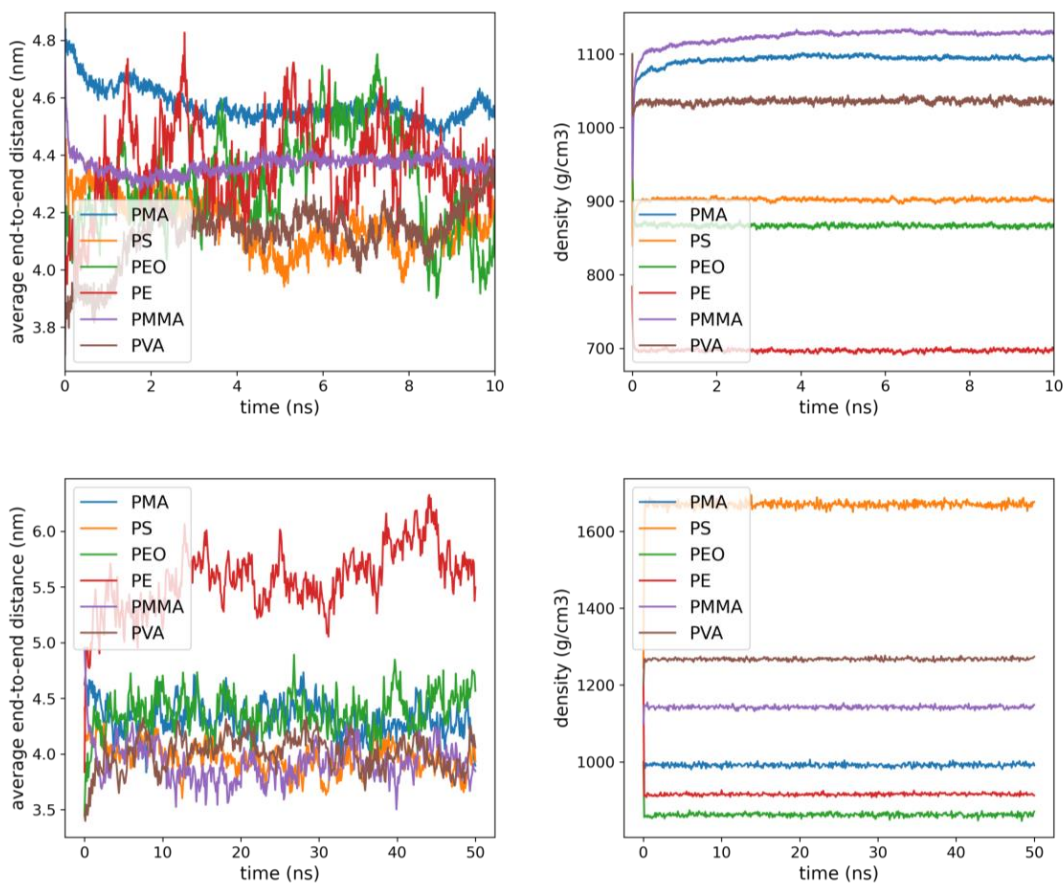

*Figure S2. Densities and end-to-end distances for equilibrations runs of melt systems at 513K. Top two panels are for GROMOS bottom two panels for Martini*

## Additional Characterization of PEO-b-PS battery

After setting up the sytem with polyply we further characterized the microphase-separated block-copolymer by computing the density profiles of the compnents along the z-axis. The result is shown in figure S3. It is clear that distinct alternating domains of PEO and PS with some degree of mixing at the interface. From the lower panel we also see that the salt is enriched inside the PEO domain relative to the edges of the PEO domain. This form of salt channel formation has been reported previously in simulation[13] and has been proposed as one possible explanation for experimental conductance trends in PEO-b-PS based batteries.[14] Furthermore the domain spacing can be computed from the total box size. After equilbriation the total box dimension in z is stable with an average value of 63nm. Considering we see three domains the average domain spacing is about 21nm in good agreement with the reported experimental value of 20nm for this systems.[15]

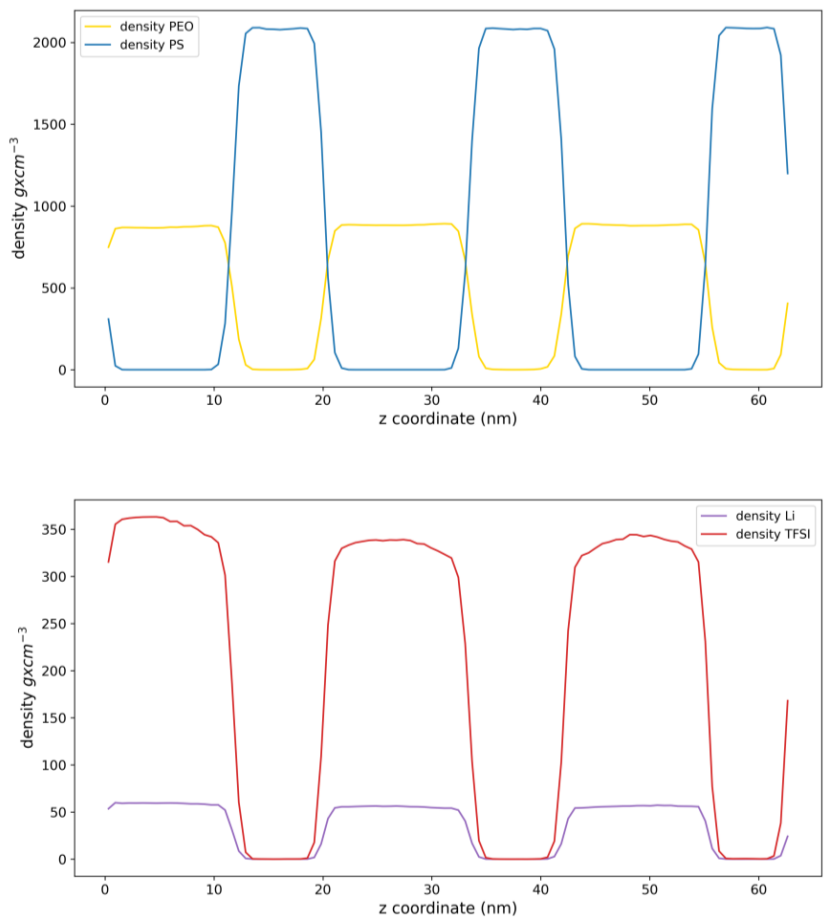

*Figure S3. Density profiles of block-copolymer LiTFSI doped battery after equilibration of 50ns;*

## Dextran molecular weight distribution

Dextran is in general a branched and polydisperse polysaccharide. The degree of branching is reasonably well characterized with literature agreeing that about 5% linkages should be present for molecular weights lower than 100,000 g/mol.[16,17] Unfortunately the polydispersity is less well characterized depends on the synthesis method and if the polymer was fractionated afterwards. As we are only interested in the polydispersity of for the sake of demonstrating polyply's capability of dealing with polydisperse polymers, we modeled the molecular weight distribution as follows: We assume that the synthesis follows the reaction kinetics of linear condensation polymers. For this type of reaction the molecular weight distribution can be computed only from knowing the extend of reaction (p in eq. 5).[2]

$$prop(N,p) = N \times p^{N-1}(1-p)^2 \quad (5)$$

Subsequently we change the extend of reaction such that the number average molecular weight becomes about 65, which is the weight used in the experimental system we aim to model.[18] The obtained polydispersity index in that case is about 1.5 which compares reasonably well to the value of 1.8 found in literature[19]. Figure S4 shows the distribution and a histogram of the 500 samples draw. Note that the smallest allowed polymer was 2 residues.

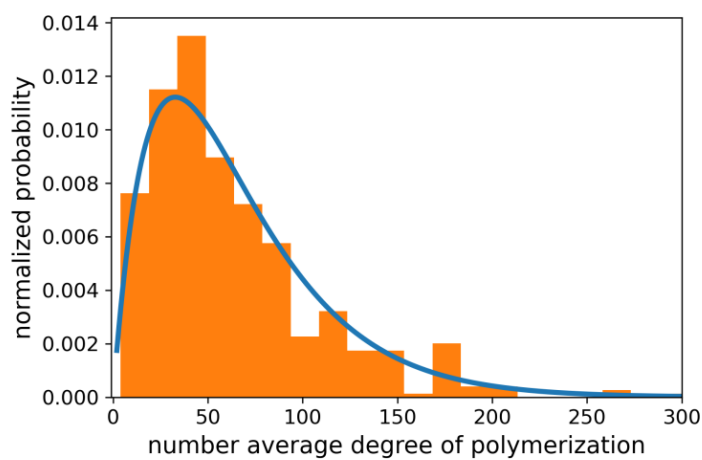

*Figure S4. Molecular weight distribution of dextran from equation x (solid line) and the samples drawn for structure generation.*